\documentclass[onecolumn,preprint,preprintnumbers,amsmath,amssymb]{revtex4}

\usepackage{graphicx}
\usepackage{dcolumn}
\usepackage{bm}

\begin{document}

\title{Power flow in a small electromagnetic energy harvesting system excited by mechanical motion}

\author{L.E. Helseth}

\affiliation{Department of Physics and Technology, University of Bergen, N-5007 Bergen, Norway}%

\begin{abstract}
In this study the power flow in a coupled mechanical and electromagnetic harvesting system in presence of both positional and electrical fluctuations is analyzed. Explicit expressions for the power into and out of the mechanical and electrical parts of the system are found in the case of weak coupling, and it is shown how the power flows between the domains consistent with energy conservation. The case of thermal fluctuations is considered in particular, and use of the fluctuation-dissipation theorem explicitly demonstrates that the power delivered to the mechanical system from the electrical system is the same as the power delivered to the electrical system from the mechanical system. On the other hand, the power dissipated in the electrical circuit is not the same as the power transferred from the mechanical domain if the electrical circuit contains its own current fluctuations. The electrical noise power dissipated in a load resistor is calculated, and found to consist of a component due to electromagnetic coupling in addition to the well-known Nyquist component. The component due to electromagnetic coupling scales with size, and becomes more important than the Nyquist component only for sufficiently small systems.
\end{abstract}

\pacs{Valid PACS appear here}
\maketitle

\section{Introduction}
Vibration energy harvesting is a method to transfer mechanical energy into electromagnetic energy with the aim of powering small-scale sensors or devices.  It is likely that future micro and nanosystems will harvest all the energy they need from the environment. Energy harvesting based on mechanical vibrations is a subject of increasing interest, due to the fact that vibrations are present in a large number of physical systems. Pioneering experimental studies on miniature vibration energy harvesting systems were reported by Yates and coworkers\cite{Williams,Shearwood}. Since then, considerable work has been done to understand theoretically, and experimentally, the extraction of electrical power from vibration harvesting systems, see Refs. \cite{Hudak,Vullers,Gammaitoni1} and references therein. In particular, considerable efforts have been concentrated on optimizing the mechanical and electrical power as well as the  electromagnetic coupling coefficient\cite{Poulin,Stephen,Cepnik,Canarella}. Studies have also shown that coupling different harvesting modes, e.g. electromagnetic and piezoelectric, may enhance the power output\cite{Challa}. Poulin et al. pointed out that electromagnetic and piezoelectric vibration harvesting systems can be modeled using the same differential equations, thus allowing a unified description of both\cite{Poulin}. Arroyo et al. compared theoretical and experimental results, demonstrating a decrease in electrical power with an increase in the electrical loss coefficient\cite{Arroyo}. Since the electrical loss coefficient, which is essentially given by the ratio between the real and imaginary part of the electrical impedance, often becomes more important as the circuit decreases, electromagnetic generators have not been miniaturized at the same pace as their piezoelectric and electrostatic counterparts. In fact, although the first electromagnetic generators were miniature devices producing low output powers, the trend has facilitated the development of much larger devices ($>cm^{3}$) producing energies far in excess of 1 mW\cite{Williams,Hudak,Vullers}. From an engineering perspective, piezoelectric and electrostatic harvesters are straightforward to fabricate down to micrometer scale, and  can in many cases be combined with existing mass-production technology solutions. On the other hand, electromagnetic harvesters are often fabricated using micromachining and manual tooling, which is irreconcilable with current wafer production technology. The belief that electromagnetic generators do not perform equally well as other types of devices when size is decreased was questioned in ref. \cite{Arroyo}, where it was found that as long as the electromagnetic coupling coefficient remains high while the electrical loss parameter stays below a certain value, there is no formal reason why electromagnetic energy harvesting systems should perform worse than their piezoelectric and electrostatic counterparts.

Until a few years ago, most of the research in the field of vibration energy harvesting concentrated on either impact or oscillation-based devices exhibiting a narrow frequency response. In many practical systems of interest it is necessary to utilize the wide-band tuning devices, and this was addressed by Sari et al., who proposed arrays of linear cantilevers each with different resonance frequency, in order to ensure that the system would be able to utilize the available vibration spectrum\cite{Sari}. Clearly, such a system may take up a larger volume than a single cantilever, and it was desirable to find methods to allow further power density optimization. In Refs. \cite{Cottone,Gammaitoni} it was demonstrated that nonlinear energy harvesting systems are not inhibited by the same bandwidth limitations as linear systems. A theoretical analysis of the electrical power output of both linear and nonlinear systems excited by random vibrations was given by Halvorsen\cite{Halvorsen}. Various approaches to improve output power, and in particular the shape of the potential well used, have been considered. In Refs. \cite{Blystad,Nguyen} the influence of an end stop to limit the motion as well as the effect of nonlinear springs in an harvester driven by colored noise was studied. Deza et al. found that the tunable Woods-Saxon oscillator may provide a useful model system when optimizing the harvesting system driven by colored noise\cite{Deza}. A survey of broadband techniques for vibration energy harvesting has been given by Twiefel and Westermann\cite{Twiefel}.

Driving an energy harvesting system by noise is not trivial, and careful analysis optimization of the amplitude and power are required. The statistical mechanics is strongly related to the much investigated topic of optimally understanding and enhancing the efficiency of Brownian motion excited by colored noise\cite{Astumian,Bouzat}. Furthermore, thermal fluctuations have been measured in superconducting and mechanical systems in different frequency bands using interferometric techniques\cite{Koch,Numata}. Theoretically, the generalized Langevin equation has been found to be able to model a range of different mechanical and electrical systems\cite{Ford,OConnell1,OConnell,Li1,Li2}, and should therefore be a very useful tool for modeling also vibration energy harvesting systems.
As the vibration energy harvesting systems become smaller, one would have to deal with thermal excitations, either as a method to enhance the performance or by reducing it using, e.g., external forces. Thermal position fluctuation reduction using controlled external dissipative forces has been pursued experimentally in the field of atomic force microscopy\cite{Liang}, but similar studies have not been undertaken for vibration harvesting systems.

The power flow in an energy harvesting system exposed to well-defined oscillating mechanical motion was analyzed in Ref. \cite{Stephen}. However, it appears that the power flow between and within the electrical and mechanical domains has not been addressed for noisy mechanical motion in such a manner that also thermal fluctuations can naturally be encountered for. Although rectification and signal conditioning of small currents is a most pressing issue in very small harvesting systems, it is also of importance to know how much power can be actually expected to flow between and in the mechanical and electrical parts of the energy harvester. In this study these questions are addressed, with particular emphasis on the power flow in presence of thermal fluctuations.

\section{The electromagnetic coupling}
Let us assume a non-relativistic system of mass m which has one degree of freedom and is described by a position operator $x(t)$. The generalized Langevin equation is given by
\begin{equation}
m\ddot{x}(t) + \int_{-\infty}^{t} \mu (t -\tau)\dot{x}(\tau)d\tau + \frac{dU(x)}{dx}  =F(t) + f_{em}(t) \,\,\, ,
\label{Langevin}
\end{equation}
where $\mu (t)$ is the memory function. The force $-dU(x)/dx$ is acting on the system, where $U(x)$ is the corresponding potential. The force $F(t)$ represents a sum of the oscillating force $f_m (t)=f_0 sin(\omega t)$ and the fluctuation force $f_f (t)$ caused by the thermally excited molecules surrounding the electromechanical system. Moreover, $f_{em} (t)$ is the electromagnetic force due to the current $i(t)$ in the electric circuit. The case where $f_{em} (t)=0$ has been analyzed in detail in Ref. \cite{Ford}, where the Langevin equation was shown to follow from a very general quantum mechanical independent oscillator model. Here we are mainly interested in the situation where there is non-negligible coupling ($f_{em} \neq 0$), with the aim of investigating the generation of electrical power in an electric circuit coupled to the mechanical system.

In order to proceed, it is useful to assume the following relationship linking current $i$ and velocity $\dot{x}$,
\begin{equation}
i(t) = i_0 (t) + \int_{0}^{\infty} \beta (\tau) \dot{x}(t-\tau) d\tau \,\,\, ,
\label{A2}
\end{equation}
where $\beta$ is a coupling coefficient associated with the coupling of mechanical into electrical energy and $i_0(t)$ is the current existing in the circuit due to thermal fluctuations. Similarly, it is possible to define another coupling coefficient $\gamma$ linking changes in position with current such that
\begin{equation}
x(t) =x_0 (t) + \int_{0}^{\infty} \gamma (\tau) i(t-\tau) d\tau \,\,\, ,
\label{A3}
\end{equation}
where $x_0(t)$ is the time-dependent position due to fluctuations or well-defined oscillations not caused by the electromagnetic coupling. We will in the following assume that the thermal current fluctuations are uncorrelated with the position fluctuations such that $<i_0 (t) \dot{x}_0(t)>=0$. In the spectral domain these equations become
\begin{equation}
i(t) =i_0 (\omega) -j\omega \beta (\omega)x(\omega)\,\,\, ,
\label{A2s}
\end{equation}
and
\begin{equation}
x(\omega) =x_0 (\omega)+ \gamma (\omega)i(\omega) \,\,\, .
\label{A3s}
\end{equation}
Here $x (\omega)$ is the Fourier transform of $x(t)$, $i (\omega)$ is the Fourier transform of $i(t)$, $\beta (\omega)=\int_{0}^{\infty} \beta (\tau) exp(j\omega \tau) d\tau$ and $\gamma (\omega)=\int_{0}^{\infty} \gamma (\tau) exp(j\omega \tau) d\tau$. By inserting Eq. \ref{A3s} into Eq. \ref{A2s} it is also possible to write the spectral current as a function of the
uncoupled spectral components $i_0 (\omega)$ and $x_0 (\omega)$
\begin{equation}
i(\omega) =\frac{i_0 (\omega) -j\omega \beta (\omega)x_0(\omega)}{1+j\omega \beta (\omega)\gamma(\omega)} \,\,\, ,
\label{A2sa}
\end{equation}
and
\begin{equation}
x(\omega) =\frac{x_0 (\omega) + \gamma (\omega)x_0(\omega)}{1+j\omega \beta (\omega)\gamma(\omega)} \,\,\, .
\label{A3sa}
\end{equation}
If $\omega |\beta (\omega)| |\gamma (\omega)| \ll 1$, one may let the denominator be unity, such that
\begin{equation}
i(\omega) \approx i_0 (\omega) -j\omega \beta (\omega)x_0(\omega) \,\,\, ,
\label{A2sb}
\end{equation}
and
\begin{equation}
x(\omega) =x_0 (\omega) + \gamma (\omega)x_0(\omega) \,\,\, .
\label{A3sb}
\end{equation}
This means that when the coupling between the electromagnetic and mechanical domain is weak, such that eqs. \ref{A2} and \ref{A3} become
\begin{equation}
i(t) = i_0 (t) + i_1 (t) \approx i_0 (t)+ \int_{0}^{\infty} \beta (\tau) \dot{x}_0 (t-\tau) d\tau \,\,\, ,
\label{A4}
\end{equation}
and
\begin{equation}
x(t)  = x_0 (t) + x_1 (t) \approx x_0 (t)+ \int_{0}^{\infty} \gamma (\tau) i_0 (t-\tau) d\tau \,\,\, .
\label{A5}
\end{equation}
Equations \ref{A4} and \ref{A5} can be viewed upon as a first-order approximation, wherein only the original position and current fluctuations contribute to the coupling, and second-order effects are neglected. Clearly, by keeping higher-order terms (i.e. not making the assumption $\omega |\beta (\omega)| |\gamma (\omega)| \ll 1$) one may also consider the case of strong coupling as well, but this is outside the scope of the current study.

Let us now link the parameter $\gamma (\omega)$ with the the spectral susceptibility of the system, since susceptibility is often used to describe a mechanical system. The susceptibility is defined through the equation $x(t)= \int_{0}^{\infty} \alpha (\tau) F(t-\tau) d\tau$, which in the frequency domain becomes $x(\omega)=\alpha(\omega)F(\omega)$ with $\alpha (\omega)=\int_{0}^{\infty} \alpha(\tau) exp(j\omega \tau) d\tau$. If $x_0=0$ and all the position fluctuations are due to applied currents in the electrical circuit, we may write Eq. \ref{A3s} as $x(\omega) = \gamma (\omega)i(\omega)$. On the other hand, we know that the Lorentz force acting on the mechanical system can be written as $f_{em} (t) =-K_m i(t)$, where $K_m$ represents the coupling between the electromagnetic and mechanical degrees of freedom. Thus, we must require $\gamma (\omega) =-K_m \alpha (\omega)$.

Similarly, it is possible to link the parameter $\beta (\omega)$ with the impedance of the electrical system. If we assume that all the current fluctuations are due to movement in the mechanical domain, we may write Eq. \ref{A2s} as $i(\omega) = -j\omega \beta (\omega)x(\omega)$. The voltage is related to the current through the impedance $Z(t)$ as $v(t)= \int_{0}^{\infty} Z(\tau) i(t-\tau) d\tau$, which in the frequency domain becomes $v(\omega)=Z(\omega)i(\omega)$ with $Z(\omega)=\int_{0}^{\infty} Z(\tau) exp(j\omega \tau) d\tau$. Thus, if the voltage spectrum $v(\omega)=-j\omega \beta (\omega)Z(\omega)x(\omega)$ is to be consistent with Faraday's law, which in the frequency domain can be written as $v(\omega)=-j\omega K_m x(\omega)$, we will have to require that $\beta (\omega)=K_m/Z(\omega)$.

\section{Power flow}
The instantaneous mechanical power due to the force $F(t)$ is defined as (in symmetrized form)
\begin{equation}
P_{M}=\frac{1}{2}\left< v(t)F(t) + F(t)v(t) \right> \,\,\, .
\label{MPower}
\end{equation}
Using Eq. \ref{Langevin}, it is found that
\begin{equation}
P_{M}=\frac{d}{dt} \left< \frac{1}{2}m\dot{x}(t) ^{2} + U(x) \right>  + P_{M0}  + P_{Mout} + P_{Min}\,\,\, ,
\label{MPower}
\end{equation}
where the first term on the left, $ d/dt <1/2m\dot{x}(t) ^{2} + U(x)> $, is zero due to the fact that the expectation value does not change with time. Moreover, $P_{M0}$ is
\begin{equation}
P_{M0}= \int _{0}^{\infty} \mu (\tau) \frac{1}{2} \left< \dot{x}(t)\dot{x}(t-\tau) + \dot{x}(t-\tau)\dot{x}(t) \right> d\tau = 2\int_{0}^{\infty}  \omega ^{2}Re\left[ \mu (\omega)\right] S_x(\omega) d\omega \,\,\, .
\label{MPowereq}
\end{equation}
Here $P_{M0}$ is the power dissipated by the lossy mechanical element (e.g. viscous forces or coupling to the radiation bath) in absence of electromagnetic coupling (as described in the previous section), and was studied in ref. \cite{Li1}. On the other hand, $P_{Mout}$ is the power transferred out of the mechanical and into the electrical domain,
\begin{equation}
P_{Mout} \approx \frac{1}{2}K_m \left< \dot{x}_0(t) i_1 (t) + i_1 (t) \dot{x}_0 (t) \right> =2K_m\int_{0}^{\infty}  \omega ^{2}Re\left[ \beta (\omega)\right] S_x(\omega) d\omega \,\,\, ,
\label{MePowerO}
\end{equation}
whereas $P_{Min}$ is the power transferred into the mechanical domain from the electrical domain
\begin{equation}
P_{Min} \approx \frac{1}{2}K_m \left< \dot{x}_1(t) i_0(t) + i_0 (t) \dot{x}_1(t) \right> =2K_m\int_{0}^{\infty} \omega Im\left[ \gamma (\omega)\right] S_i(\omega) d\omega \,\,\, .
\label{MePowerI}
\end{equation}
In the equations above $S_x(\omega)=<x_0 ^{2}(\omega)>$ is the position spectral density and $S_i(\omega)=<i_0 ^{2}(\omega)>$ is the current spectral density, both due to the position and current fluctuations not induced by the electromagnetic coupling (hence the subscript). We have used the first-order approximation discussed in the previous section to arrive at Eqs. \ref{MePowerO} and \ref{MePowerI}, which therefore are valid for weak coupling ($\omega |\beta (\omega)| |\gamma (\omega)| \ll 1$).

In general, the model system here allows for $P_{Mout} + P_{Min} \neq 0$, which means that power is either transferred into or out of the mechanical domain, with corresponding impact on the environment. However, in the special case that the system is in equilibrium, we must require $P_{Mout} + P_{Min} = 0$ such that the power transferred out of the mechanical domain into the electrical domain is the same as that coming into the mechanical from the electrical domain. Comparing Eqs. \ref{MePowerO} and \ref{MePowerI} we then have $\omega Re\left[ \beta (\omega)\right] S_x(\omega)=-Im\left[ \gamma (\omega)\right] S_i(\omega)$. That this condition is fulfilled for a system near equilibrium can be seen by applying the fluctuation-dissipation theorem of Callen and Welton, which states that $S_x(\omega)=\frac{\hbar}{2\pi} Im\left[ \alpha(\omega)  \right] coth\left ( \frac{\hbar \omega}{2k_B T}\right) $ and $S_i(\omega)=\frac{\hbar}{2\pi} \omega Re\left[ Z(\omega)\right] |Z(\omega)| ^{-2} coth\left ( \frac{\hbar \omega}{2k_B T}\right)$ \cite{Callen}. Here $k_B$ is Boltzmann's constant and $h=2\pi \hbar$ is Planck's constant. Thus, near equilibrium the net dissipated mechanical power is given by Eq. \ref{MPowereq}.

The total power in the electrical system is composed of one component flowing into the mechanical domain and another being dissipated in the electrical circuits. The component flowing into the mechanical domain is, due to energy conservation, given by Eq. \ref{MePowerI}, and is not of particular interest here. Of more interest is it to find the power dissipated in the electrical circuit, since that power can be utilized in the electrical energy harvesting system. The average dissipated electrical power is given by $P_{E}=(1/2)<vi+iv>$, where the voltage is related to the current through the impedance $Z(t)$ as $v(t)= \int_{0}^{\infty} Z(\tau) i(t-\tau) d\tau$, which in the frequency domain becomes $v(\omega)=Z(\omega)i(\omega)$ with $Z(\omega)=\int_{0}^{\infty} Z(\tau) exp(j\omega \tau) d\tau$. The total power dissipated in the electrical circuit is given by
\begin{equation}
P_{E}= \int_{-\infty}^{\infty}  Z(-\omega)  <i^{2}(\omega)>  d\omega = P_{E0} + P_{E1}\,\,\, ,
\label{EPowere1}
\end{equation}
where
\begin{equation}
P_{E0} \approx 2\int_{0}^{\infty}  Re\left[ Z(\omega)\right]  S_i(\omega)  d\omega \,\,\, ,
\label{EPowere1a}
\end{equation}
is due to thermal current fluctuations independent of the electromagnetic coupling, and
\begin{equation}
P_{E1} \approx 2\int_{0}^{\infty}  Re\left[ Z(\omega)\right]  \omega ^{2} |\beta(\omega)|^{2} S_x(\omega)  d\omega \,\,\, .
\label{EPowere1b}
\end{equation}
is the power flowing into the electrical circuit from the mechanical system. To obtain Eqs. \ref{EPowere1a} and \ref{EPowere1b} we have used the first order approximation discussed in the previous section.

By noting that $Re\left[ \beta (\omega)\right] = Re\left[ Z (\omega)\right]/|Z(\omega)|^{-2}$, it is seen that $P_{E1}=P_{Mout}$. Thus, the power transferred out of the mechanical domain is dissipated in the electrical circuit, and energy conservation is ensured in the model used here. It is interesting to note that the power dissipated in the electrical circuit is not the same as the power transferred from the mechanical domain. One may define a ratio $\eta _{EM} =P_{E}/P_{Mout}$, which becomes
\begin{equation}
\eta _{EM} \approx 1+ \frac{\int_{0}^{\infty}  Re\left[ Z(\omega)\right]  S_i(\omega)  d\omega}{K_m\int_{0}^{\infty}  \omega ^{2}Re\left[ \beta (\omega)\right] S_x(\omega) d\omega} \,\,\, ,
\label{ratioeta}
\end{equation}
where we have used that $P_{E1}=P_{Mout}$. It is seen that for a system in thermal equilibrium that $\eta _{EM} >1$, where the additional (noise) power dissipated in the electrical circuit occurs as a result of the thermal fluctuations. In general, one needs to know the spectral densities ($S_i(\omega)$ and $S_x(\omega)$), the impedance $Z(\omega)$ and the susceptibility $\alpha (\omega)$  in order to compute the dissipated power in an electrical circuit using the formalism presented above. We will in the next section look at a simple example which allows us to connect the theory presented here with some special cases previously encountered in the literature.

\section{Electrical power dissipation in presence of a resistive load}
Let us consider a harvesting system where there is an electromechanical coupling between the electromagnetic and the mechanical systems, and where the electrical circuit has an internal resistance $R_i$ connected to a resistive load $R_L$. The friction coefficient is constant such that $Re\left[ \mu (\omega) \right]=c$. An estimate suggests that in order to use the first-order approximation of equations \ref{A4} and \ref{A5} one needs to set $K_m ^{2}/(R_L +R_i)c \ll 1$. As pointed out by Weber, equivalent circuits are not always good models of physically reliable circuits under all conditions\cite{Weber}. Deviations are in particular expected at high frequencies, i.e. when the smallest wavelength of the electromagnetic waves becomes comparable to the size of the circuit. However, it should be emphasized that due to the natural frequency filtering effect of the mechanical oscillator, such high frequencies are ruled out in basically all practical systems. We will also neglect radiation reaction forces associated with self-induced currents as studied in Ref. \cite{Blanco} since these only become important at very high frequencies.

For the moment we assume that the mechanical system is forced by an external force $f_m \gg f_{em}$ and $f_m \gg f_f$. The system oscillates sinusoidally at the resonance frequency $\omega _0$ such that
$S_x (\omega) =(x_0 ^{2}/4)\delta (\omega - \omega _0)$.  Here we would like to evaluate the electrical power dissipated in the load resistor $R_L$, which is different from the total dissipated power given by Eq. \ref{EPowere1}. To maximize the power transfer, one requires impedance matching with $R_L=R_i$ as the resistance of the load. The power to be evaluated is now given by $P_{Eload}=<v_L i>$, where $v_L=R_L i$, such that the power dissipated in the load can be found to be
\begin{equation}
P_{ELoad} \approx  \frac{K_m ^{2}}{4R_L } \frac{\omega _0 ^{2}x_0 ^{2}}{2}  + k_B T \Delta f \,\,\, .
\label{EPower4a}
\end{equation}
Here $\Delta f$ is the frequency bandwidth of the electrical circuit. It should be emphasized that the first part of Eq. \ref{EPower4a} is the same as Eq. 47 in Ref. \cite{Stephen} for an impedance-matched load. The second part of Eq. \ref{EPower4a} is just the noise power associated with thermal fluctuations in the electrical circuit, as was found by Nyquist\cite{Nyquist}. Note that both the internal resistance and the load resistor generate equal amounts of thermal noise power, such that in thermal equilibrium there is no net power flow in the circuit. It is seen that in the simple approximation considered here the total power dissipated in the electrical load is just the sum of the special cases considered in previous studies. Only in the case that the vibration amplitude is very small, $x_0 \sim \sqrt{8k_B T \Delta f R_L }/(\omega _0 x_0 K_m)$, the Nyquist noise power becomes comparable to the power transferred due to electromagnetic coupling. However, when the latter term becomes this small, it would be more correct to look at the spectral densities caused by thermal fluctuations, both for position and current.

To this end, we now set $f_m =0$ and consider a thermally fluctuating system where $S_x(\omega)=\frac{\hbar}{2\pi} Im\left[ \alpha(\omega)  \right] coth\left ( \frac{\hbar \omega}{2k_B T}\right) $ and $S_i(\omega)=\frac{\hbar}{2\pi} \omega Re\left[ Z(\omega)\right] |Z(\omega)| ^{-2} coth\left ( \frac{\hbar \omega}{2k_B T}\right)$. Furthermore, we assume that $\hbar \omega /k_B T \ll 1$ and that the electrical system has an upper frequency $\Delta \omega =2\pi \Delta f $. Under such circumstances the power dissipated in the load can be found as
\begin{equation}
P_{ELoad} \approx P_{Etransfer} + P_{Ethermal} =  \frac{K_m ^{2}}{4R_L} \frac{k_B T}{m} + k_B T \Delta f \,\,\, .
\label{EPower6bb}
\end{equation}

\begin{figure}
\includegraphics[width=10cm]{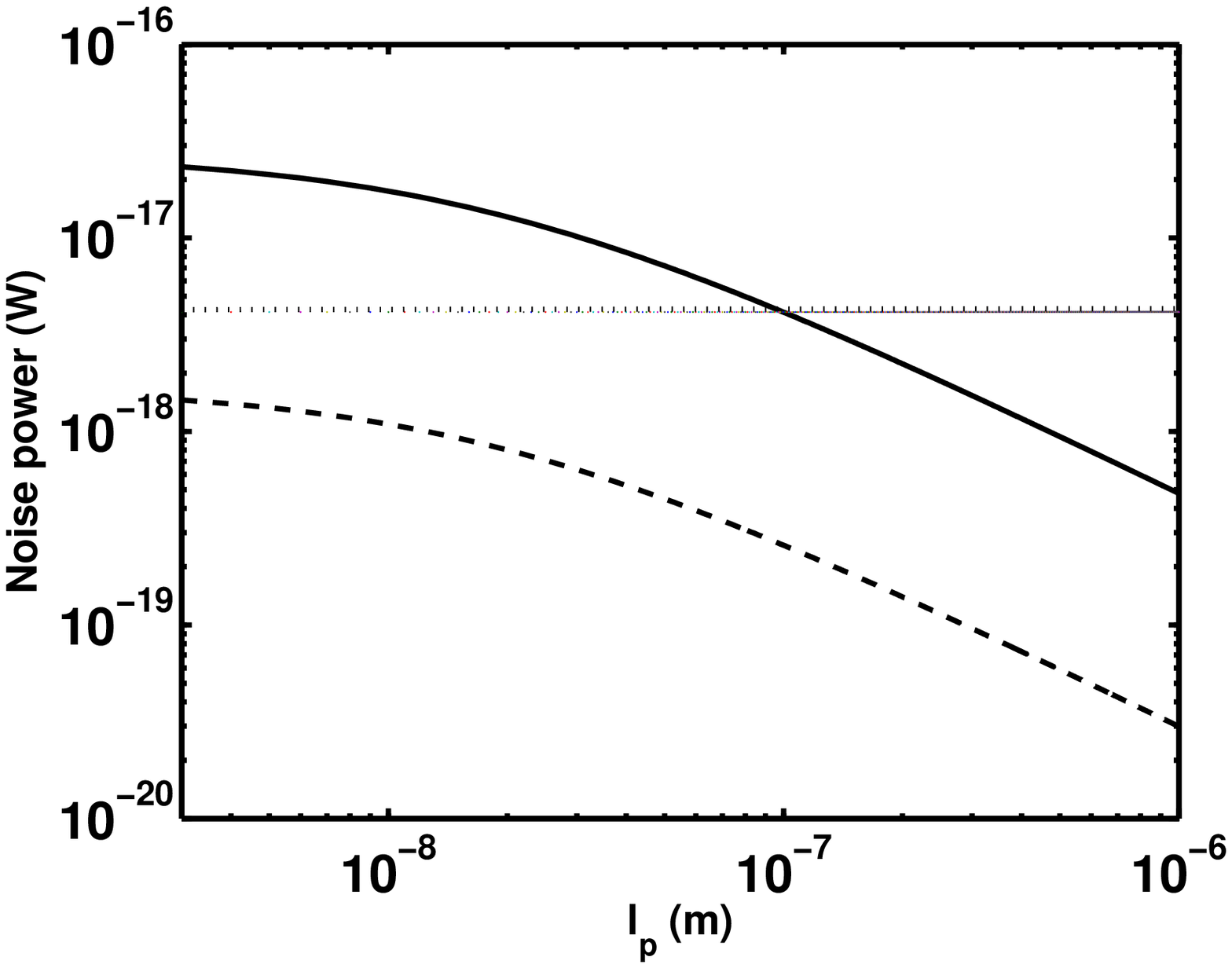}
\caption{\label{f1} The electrical noise power dissipated in the load resistor as a function of system size. See text for details.}
\vspace{2cm}
\end{figure}

For a macroscopic system the thermal fluctuations are negligible, and the noise power predicted by Eq. \ref{EPower6bb} is under normal circumstances indeed orders of magnitude smaller than the power generated in the energy harvesting system reported in e.g. Refs. \cite{Williams,Shearwood}. However, when the system becomes smaller, approaching the micro and nanometer scale, one expects the noise power to become increasingly important. Here it should be emphasized that, unlike the Nyquist noise power, the transferred power from the mechanical domain scales inversely with the system size. To see an example of this, assume that the harvesting system is composed of a cube of sides $l_p$, and that a conductor of length $4l_p$, width $l_p$ and thickness $t$ (where $t\ll l_p$) is wrapped around it such that it makes one turn. In order to obtain an analytical expression, we will now assume that $K_m \approx 4Bl_p$. Although the exact field distribution and geometry is not accounted for in such a simple expression, it does give a correct order of magnitude, and is therefore a useful first approach. The conductor acts like a thin shell covering four sides of the cube, similar to that of a colloid covered by a metallic shell. The technical details of connecting the conductor with the external load $R_L$ will not be considered here. The mass of the system is $m=\rho _m l_p ^{3} + 4\rho _c l^{2} t$, where $\rho _m$ is the density of the cube and $\rho _c$ the density of the conductor. The conductor, with its conductivity $\rho _e$ and square cross section $A=l_p t$, exhibits an intrinsic resistance $R_i =4\rho _e l_p /A$. Now the noise power transferred from the mechanical to the electrical domain is expressed as
\begin{equation}
P_{Etransfer} \approx \frac{B^{2} t k_B T}{\rho _e  \left( \rho _m l_p + 4\rho _c t \right)  } \,\,\, .
\label{EPower6ab}
\end{equation}
Figure \ref{f1} shows this noise power when $k_B=1.38 \times 10^{-23}$ $JK^{-1}$, $T=300$ K, $\rho _m =2000$ $kgm^{-3}$, $\rho _c =8960$ $kgm^{-3}$, $\rho _e =1.7 \times 10^{-8}$ $\Omega m$ and $t=1$ nm. The solid line corresponds to $B=2$ T and the dashed line to $B=0.5$ T. Notice that the power appears to flatten out below $l_p \sim 10^{-8}$ m due to the fact that the thickness of the conductor has been assumed to be constant. However, for $l_p \gg 10^{-8}$ m the power decreases roughly as $l_p ^{-1}$. Other selections of which dimensions to hold constant and which to scale with system size may give rise to different power scaling, but the order of magnitude is not expected to change. It is seen that for the model given here powers less than $10^{-16}$ W can be expected.
The dashed line of Fig. \ref{f1} shows the Nyquist noise power ($P_{Ethermal} \approx k_B T \Delta f $) assuming $\Delta f= 1000$ Hz, for comparison. Note that only for nanometer-sized systems the transferred mechanical noise (due to the electromagnetic coupling) power exceeds the Nyquist noise power for the parameters used here.

\section{Conclusion}
The power flow in a coupled electromechanical system has been analyzed. The terms identifying the power into and out of the mechanical and electrical parts of the system were found, and it was shown explicitly that the mechanical power transferred out of the mechanical domain was dissipated in the electrical circuit. In addition, the electrical circuit also dissipated Nyquist noise power due to thermal fluctuations, thus causing the total dissipated electrical power to be larger than the mechanical power fed into the electrical circuit. The dissipated electrical power was analyzed in the case of simple resistive impedance-matched load, and it was found how the transferred mechanical power scaled with system size. However, the formalism presented here allows one to compute the power dissipated in the mechanical and electrical circuits for other systems following the generalized Langevin equation as long as the position and current spectral densities, the electrical impedance and the mechanical susceptibility can be determined.

\newpage

\newpage


\begin{references}

\bibitem{Williams} C.B. Williams and R.B. Yates,
{\em Sens. Act. A, $\bf{52}$}, 8-11 (1996).

\bibitem{Shearwood} C. Shearwood and R.B. Yates,
{\em Electronics Letters, $\bf{33}$}, 1883 (1997).

\bibitem{Hudak} N.S. Hudak and G.G. Armatucci
{\em J. Appl. Phys., $\bf{103}$}, 101301 (2008).

\bibitem{Gammaitoni1} L. Gammaitoni,
{\em Contemp. Phys., $\bf{53}$}, 119 (2012).

\bibitem{Vullers} R.J.M. Vullers, R. van Schaijk, I. Doms, C. Van Hoof and R. Mertens,
{\em Solid-State Electronics, $\bf{53}$}, 684–693 (2009).

\bibitem{Stephen} N.G. Stephen,
{\em J. Sound and Vibration, $\bf{293}$}, 409-425 (2006).

\bibitem{Poulin} G. Poulin, E. Sarraute, F. Costa,
{\em Sens. Act. A, $\bf{116}$}, 461 (2004).

\bibitem{Cepnik} C. Cepnik, O. Radler, S. Rosenbaum, T. Strohla and U. Wallrabe,
{\em Sens. Act. A, $\bf{167}$}, 416-421 (2011).

\bibitem{Canarella} J. Canarella, J. Selvaggi, S. Salon, J. Tichy and D.A. Borca-Tasciuc,
{\em IEEE Trans. Mag., $\bf{47}$}, 2076-280 (2011).

\bibitem{Challa} V.R. Challa, M.G. Prasad and F.T. Fisher,
{\em Smart Mater. Struct., $\bf{18}$}, 095029 (2009).

\bibitem{Arroyo} E. Arroyo, A. Badel, F. Formosa, Y. Wu and J. Qiu,
{\em Sens. Act. A, $\bf{183}$}, 148 (2012).

\bibitem{Sari} I. Sari, T. Balkan and H. Kulah,
{\em Sens. Act. A, $\bf{145-146}$}, 405 (2008).

\bibitem{Cottone} F. Cottone, H. Vocca and L. Gammaitoni,
{\em Phys. Rev. Lett., $\bf{102}$}, 080601 (2009).

\bibitem{Gammaitoni} L. Gammaitoni, I. Neri, and H. Vocca,
{\em Appl. Phys. Lett., $\bf{94}$}, 164102 (2009).

\bibitem{Halvorsen} E. Halvorsen,
{\em J. Microelectromech. Syst., $\bf{17}$}, 1061–71 (2008).

\bibitem{Blystad} L.C.J. Blystad and E. Halvorsen,
{\em Smart Mater. Struct., $\bf{20}$}, 025011 (2009).

\bibitem{Nguyen} S.D. Nguyen and E. Halvorsen,
{\em J. Microelectromech. Syst., $\bf{20}$}, 1225 (2011).

\bibitem{Deza} J. Ignacio Deza, R. R. Deza and H.S. Wio,
{\em Europhys. Lett., $\bf{100}$}, 38001 (2012).

\bibitem{Twiefel} J. Twiefel and H. Westermann,
{\em J. Intelligent Material Systems and Structures, $\bf{24}$}, 1291-1302 (2013).

\bibitem{Astumian} R.D. Astumian and P. H\"{a}nggi,
{\em Physics Today, $\bf{55}$}, 33 (2002).

\bibitem{Bouzat} S. Bouzat and H.S. Wio,
{\em E. Phys. J. B, $\bf{41}$}, 97 (2004).

\bibitem{Koch} R.H. Koch, D.J. van Harlingen and J. Clarke,
{\em Phys. Rev. B, $\bf{26}$}, 74 (1982).

\bibitem{Numata}  K. Numata, M. Ando, K. Yamamoto, S. Otsuka, and K. Tsubono,
{\em Phys. Rev. Lett., $\bf{91}$}, 260602 (2003).

\bibitem{Ford} G.W. Ford, J.T. Lewis and R.F. O'Connell,
{\em Phys. Rev. A, $\bf{37}$}, 4419 (1988).

\bibitem{OConnell1}  R.F. O'Connell,
{\em Phys. Rev. D, $\bf{64}$}, 022003 (2001).

\bibitem{OConnell} R.F. O'Connell and J. Zuo,
{\em Phys. Rev. A, $\bf{67}$}, 062107 (2003).

\bibitem{Li1} X.L. Li, G.W. Ford and R.F. O'Connell,
{\em Phys. Rev. E, $\bf{48}$}, 1547 (1993).

\bibitem{Li2} X.L. Li, G.W. Ford and R.F. O'Connell,
{\em Physica A, $\bf{193}$}, 575 (1995).

\bibitem{Liang} S. Liang, D. Medich, D.M. Czajkowsky, S. Sheng, J.Y. Yuan and Z. Shao,
{\em Ultramicroscopy, $\bf{84}$}, 119-125 (2000).

\bibitem{Callen}  H.B. Callen and T.A Welton,
{\em Phys. Rev., $\bf{83}$}, 34 (1951).

\bibitem{Weber}  J. Weber,
{\em Phys. Rev., $\bf{101}$}, 1620-1626 (1956).

\bibitem{Blanco}  R. Blanco, H.M. Franca, E. Santos, R.C. Sponchiado,
{\em Phys. Lett. A, $\bf{282}$}, 349-356 (2001).

\bibitem{Nyquist}  H. Nyquist,
{\em Phys. Rev., $\bf{32}$}, 110 (1928).




\end{references}
\end{document}